\documentclass{sig-alternate-05-2015}

\def\sharedaffiliation{%
\end{tabular}
\begin{tabular}{c}}
\usepackage{amssymb}
\usepackage{graphicx}
\usepackage[ruled]{algorithm2e}
\usepackage{float}
\usepackage{color}
\usepackage{soul}
\usepackage{amsmath}
\begin{document}

\setcopyright{acmcopyright}

\CopyrightYear{2017} 
\setcopyright{rightsretained} 
\conferenceinfo{MobiQuitous 2017}{November 7--10, 2017, Melbourne, VIC, Australia}\isbn{978-1-4503-5368-7/17/11}
\doi{https://doi.org/10.1145/3144457.3144506}

%


\title{Identifying Recent Behavioral Data Length in Mobile Phone Log}

\numberofauthors{3}
    \author{
    \sharedaffiliation
      \alignauthor Iqbal H. Sarker\\ 
      \affaddr{Dept. of Computer Science and Software Engineering, Swinburne University of Technology, VIC, Australia.}\\
      \alignauthor Muhammad Ashad Kabir\\    
       \affaddr{School of Computing and Mathematics, Charles Sturt University, NSW, Australia.}\\
      \alignauthor Alan Colman, Jun Han\\ 
            \affaddr{Dept. of Computer Science and Software Engineering, Swinburne University of Technology, VIC, Australia.}\\
          }

\maketitle
\begin{abstract}
Mobile phone log data (e.g., phone call log) is not static as it is progressively added to day-by-day according to individual's diverse behaviors with mobile phones. Since human behavior changes over time, the most \textit{recent pattern} is more interesting and significant than older ones for predicting individual's behavior. The goal of this poster paper is to identify the recent behavioral \textit{data length} dynamically from the entire phone log for recency-based behavior modeling. To the best of our knowledge, this is the first \textit{dynamic} recent log-based study that takes into account individual's \textit{recent behavioral patterns} for modeling their phone call behaviors.
\end{abstract}

%
%


%
%

%
%



\section{Introduction}
Nowadays, mobile phones have become part of our life. Mobile phones can record various types of contextual data related to a user's phone call activities. In the real life, individual mobile phone users may behave differently (Accept $|$ Reject $|$ Missed $|$ Outgoing) in different contexts such as time-of-the-day, days-of-the-week, social situations (e.g., meeting), locations, social relationships (e.g., mother) between individuals \cite{sarker2017anapproach}. Modeling an individual's phone call behavior utilizing \textit{recent log data} may assist them in their daily activities through capabilities, such as call interruption management, call firewall, call reminder systems. 

Since human behavior changes over time, the most \textit{recent pattern} is more interesting and significant than older ones for predicting individual's phone call behavior. To model users' present behavior, a number of research \cite{lee2010adaptive, phithakkitnukoon2011behavior} have used the behavioral patterns of recent mobile phone log, e.g., last 3 months data, to predict the future behavior rather than the patterns derived from the entire historical logs. However, such arbitrary \textit{static} data length is not meaningful to predict future phone call behavior as the time frame of the data length depends on when in the recent past behavior of a user has been changed significantly. Assume that as per log data the user has a call `reject' behavioral pattern on Monday[10:00AM-12:00PM] as she used to have a regular meeting at that time. Currently, she has no meeting at that time period on Monday and she typically `accepts' incoming phone calls. So, for this example, the past `reject' behavioral pattern, even with high evidence according to log data, is not appropriate to predict her future behavior. Therefore, \textit{identifying the  optimal length} of log data that encompasses \textit{recent behavioral patterns} is the key for modeling individual's recency-based behavior.

To identify such data length, if we consider only a short length (e.g., last week's data) as indicative of recent behavior, there may not be enough data instances in that time period to infer a valid rule for predicting future phone call behavior. Creating rules based on observations with so little ``support'' is unlikely to be effective \cite{sarker2017individualized}. On the other hand, if we take into account comparatively longer lengths (e.g., last 6 months data) as indicative of recent behavior, we could get greater \textit{support} but it might result a greater \textit{behavioral variations} thus decrease the confidence of some expected rules. As a consequence, we may miss these rules because of low confidence. Therefore, the \textit{optimal length} of log data that reflects the recent behavior of an individual needs to be identified for recency-based behavioral modeling.

\section{Our Approach}
\begin{figure}[h]
	\centering
	\includegraphics[width=\linewidth, height=2.3cm]{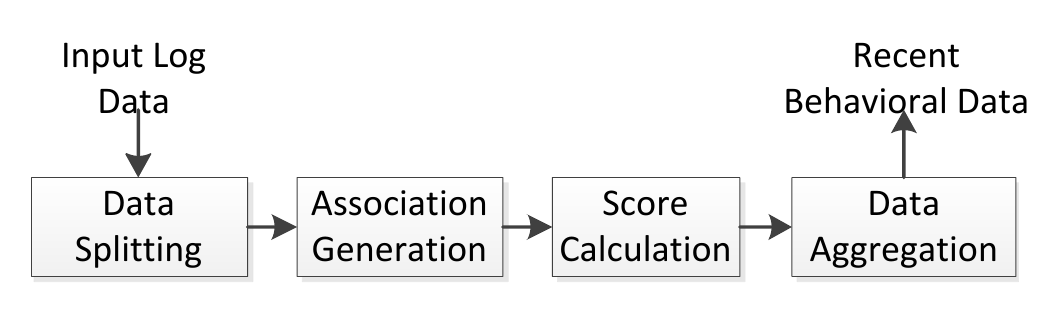}
	\caption{An overview of identifying recent behavioral log data}
	\label{fig:overview}
\end{figure}

This section presents our optimal log data length identification approach in step by step. Figure \ref{fig:overview} shows an overview of our approach.

\textbf{Step1 (Data Splitting)}: In this step, we split the entire log into week-wise data as the time-of-the-week is the most important aspect of user behavior in a mobile-Internet portal \cite{halvey2006time}. We choose weekly basis splitting because individuals' behaviors are not identical for all days in a week (Monday, Tuesday,..., Sunday), may differ from day-to-day that is captured in a week.  

\textbf{Step2 (Association Generation)}: In this step, for each week-wise data we generate context associations (combination of contexts in datasets) by adding the contexts incrementally according to the precedence of contexts. To identify the precedence of contexts in a dataset, we calculate information gain which is a statistical property that measures how well a given attribute separates training examples into targeted behavior classes \cite{sarker2017anapproach}. The one with the highest information is considered as the highest precedence context.

\textbf{Step3 (Score Calculation)}: Once we have generated the context associations, we then calculate the conflict score based on conflict behavior for each association between two adjacent weeks. For this, we first identify the \textit{dominant behavior} (maximum number
of occurrences) \cite{sarker2017individualized}, as we do not expect always 100\% like behavior of a user for a particular association. For instance, say a user 85\% rejects, 10\% accepts and 5\% misses the incoming calls for a particular association of context (e.g., meeting, office), then `reject' will be the dominant behavior for that association. We start scanning from the most recent week $[n]$ and continue to all previous weeks $[n$-$1, n$-$2, ..., 1]$ one by one until getting significant behavioral variations for the generated associations between the weeks.

\textbf{Step4 (Data Aggregation)}: In this final step, we aggregate the week-wise data based on similar behavioral patterns identified by conflict score. For identifying behavioral similarity, we calculate conflict score rather than likelihood as we do not expect similar contexts in each week. If the conflict score of two adjacent weeks is 0\% (no conflict), the behavioral patterns are highly similar in these two weeks. If there is a significant change in the conflict score of two adjacent weeks, we stop scanning and set a boundary line for recent similar behavioral patterns. In this way, for some users, recent behavioral patterns are found by aggregating large number of weeks and for some users a smaller number of weeks depending on how the user's behavior changes over time-of-the-week in different contexts.

\section{Experiments}
We have conducted experiments on real phone log datasets of two individual mobile phone users (randomly selected from Massachusetts Institute of Technology (MIT) Reality Mining dataset~\cite{eagle2006infering}). We extract 7-tuple information of the call record for each phone user from the datasets: {date of call, time of call, call-type, call duration, location, relationship, call ID}. These datasets contain three types of phone call behavior, e.g., incoming, missed and outgoing. As can be seen, the users' behaviors in accepting and rejecting calls are not directly distinguishable in incoming calls in the dataset. As such, we derive accept and reject calls by using the call duration. If the call duration is greater than 0 then the call has been accepted; if it is equal to 0 then the call has been rejected \cite{sarker2017individualized}. We also pre-process the time-series data in mobile phone log as it is continuous and numeric. For this, we use BOTS technique \cite{sarker2017individualized} for producing behavior-oriented time segments.

\begin{figure}[h]
	\centering
	\includegraphics[width=\linewidth, height=1.7cm]{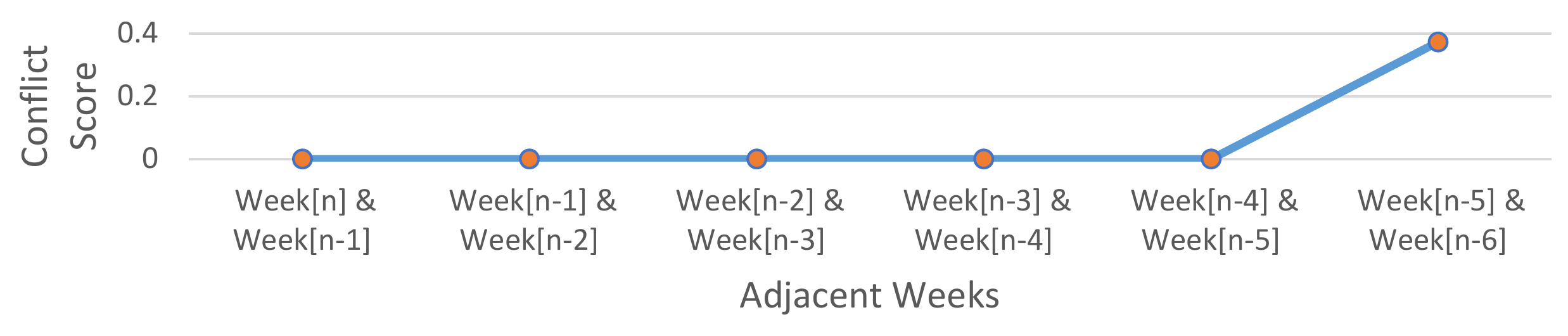}
	\caption{Conflict score for User 1}
	\label{fig:user-RM-03}
\end{figure} 
\begin{figure}[h]
	\centering
	\includegraphics[width=\linewidth, height=1.7cm]{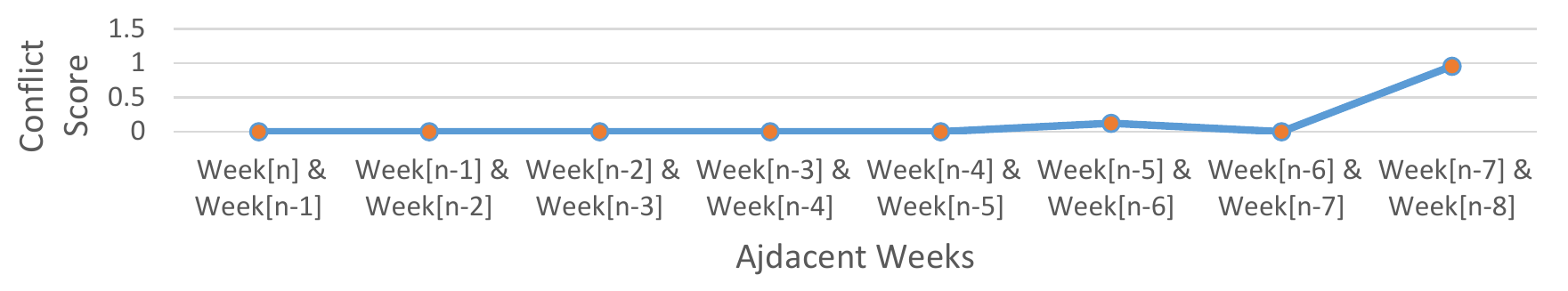}
	\caption{Conflict score for User 2}
	\label{fig:user-RM-97}
\end{figure} 

Figure \ref{fig:user-RM-03} and Figure \ref{fig:user-RM-97} show the conflict score between two adjacent weeks starting from most recent week (week n) for User 1 and User 2 respectively. From Figure \ref{fig:user-RM-03} we found that the behavioral patterns are similar from week n to week[n-5] and a significant variation has been occurred between week[n-5] and week[n-6] for User 1. In other words, the last 6 weeks data is the recent log data that represents the recent behavioral patterns of User 1. Similarly, Figure \ref{fig:user-RM-97} shows that the last 8 weeks data is the recent log data that represents the recent behavioral patterns of User 2. As the behaviors of different individuals are not identical in the real word, such data length may differ from user-to-user according to their unique behavioral patterns.  

\section{Conclusion}
In this paper, we have presented a novel approach to identify recent behavioral data length in mobile phone log dynamically by analyzing individual's behavioral patterns. Although we choose the phone call behavior as an example, our approach is also applicable to other application domains. We believe that our concept of dynamic recent log data helps both the researchers and application developers for predicting behavior of end mobile phone users according to their needs in various real-world applications, such as intelligent call interruption handling, mobile notification management systems, mobile application recommender systems etc.

\bibliographystyle{plain}
\bibliography{bibfile/recency}

\end{document}